\documentstyle[preprint,revtex]{aps}

\begin{document}
\thispagestyle{empty}


\begin{center}
\hfill{IP-ASTP-07-94}\\
\hfill{May 1994}

\vspace{1 cm}

\begin{title}
Cosmological Evolution of Scale-Invariant Gravity Waves
\end{title}
\vspace{1 cm}

\author{Kin-Wang Ng and A.~D.~Speliotopoulos}
\vspace{0.5 cm}

\begin{instit}
Institute of Physics, Academia Sinica\\ Taipei, Taiwan 115, R.O.C.
\end{instit}
\end{center}
\vspace{0.5 cm}

\begin{abstract}
The evolution of scale-invariant gravity waves from the early universe
is analyzed using an equation of state which smoothly interpolates
between the radiation dominated era and the present matter dominated
era. We find that for large wavenumbers the standard
scale-invarant wavefunction for the gravity wave
severely {\it underestimates\/} the actual size of the gravity wave.
Moreover, there is a definite shift in the {\it phase\/} of the
gravity wave as it crosses the radiation-matter phase transition.
The tensor-induced anisotropy of the cosmic
microwave background and the present spectral energy
density of the gravity wave is then calculated using these results.

\vspace{1 cm}
\end{abstract}
\newpage

It is well known that a scale-invariant (SI) spectrum for gravity waves (GW)
can be produced in an inflationary cosmology \cite{star}. During inflation,
fluctuations in the gravitational field will be red-shifted
out of the horizon, after which they freeze and remain at
a constant amplitude. Much later when they re-enter the
horizon during either the radiation-dominated (RD) or matter-dominated
(MD) era, they will appear as classical GW. Via the Sachs-Wolfe
effect \cite{sach}, these primordial GW will leave imprint on the
cosmic microwave background (CMB) \cite{ruba}. In fact, recent
measurements of the power spectrum extracted from the
large-angular-scale anisotropy of CMB \cite{smoo} suggests a behavior
that may be explained by these primordial GW \cite{krau,crit}.
Moreover, the direct detection of GW is one of the main goal of many
future experiments. Tracing the precise evolution of these GW is
therefore mandatory and important.

In this Letter we shall analyse in detail the time evolution of a SI
spectrum of GW in the standard hot big-bang cosmology. This subject
has been extensively studied before, but in all previous approaches one either
considers only the long wavelength behavior of the GW, and/or a
sudden change of equation of state as the universe moves from the RD
to MD eras is adopted. Here the novelty is that we shall
include a smooth radiation-matter phase transition, which will allow us to
obtain specific solutions of the wave equation across the transition.
By doing so, we shall find a spectrum for the short-wavelength GW that
is radically different from what is currently accepted. Moreover, we
shall be able to explicitly determine the phase of the GW.
We then apply our result in the calculation of the tensor-induced
anisotropy of CMB, and the present spectral energy density of GW.
Attempts at incorporating a smooth RD to MD transition in the
equation of state has also been made by using transfer function
methods \cite{TWL}.

We assume a spatially flat two-component universe containing
radiation and dust. The metric that we will use throughout is
the flat Robertson-Walker metric: $ds^2 = dt^2-a^2(t) d{\bf x}^2 =
a^2(\eta') ( d\eta'^2-d{\bf x}^2)$ where $a(\eta')$ and
$d\eta'=dt/a(t)$ are the scale factor and conformal time
respectively. We will consider a monochromatic plane GW of wavenumber
$k'$. We first define the new variables $\eta\equiv (\sqrt 2
-1)\eta'/\eta'_{eq}$, $k\equiv k'\eta'_{eq}/(\sqrt 2 -1)$.
$\eta'_{eq}$ is the time at which the energy density of radiation is
equal to that of matter. The equation of motion of the GW amplitude
is then given by
\cite{gris}
\begin{equation}
\frac{d^2h}{d\eta^2} +  \frac{2}{a} \frac{da}{d\eta}
\frac{dh}{d\eta} + k^2h =0.
\label{e4}
\end{equation}
In a generic inflationary model,
a SI spectrum of GW can cover a very wide range of $k$ (for instance,
$10^{-5}<k<10^{23}$) \cite{kolb}. For these $k$'s the standard
boundary conditions for $h(\eta)$ are
\begin{equation}
h(\eta)=1\qquad{\rm and}\qquad
\frac{dh(\eta)}{d\eta}=0\qquad{\rm as}\qquad
\eta\to 0.
\label{e5}
\end{equation}

In the standard solution of a SI spectrum, which we shall call the SI
solutions, Eq.~$(\ref{e4})$ is solved in the RD and MD universes
seperately \cite{ruba} giving
\begin{equation}
h_{rm}(\eta) = \cases{j_0(k\eta), &if $\eta\ll\eta_{eq}$;\cr
                 3j_1(k\eta)/(k\eta), &if $\eta\gg\eta_{eq}$,\cr}
\label{e6.1}
\end{equation}
where $j_\nu (k\eta)$ and $y_\nu (k\eta)$ (below) are the spherical Bessel
functions of order $\nu$ of the first and second kind, respectively.

As a first attempt to take into account the radiation-matter phase transition
in the analysis of the evolution of GW, one typically makes the
sudden approximation
\begin{equation}
\frac {2}{a} \frac {da}{d\eta} =\cases{2/\eta, &if $\eta<\eta_{eq}$;\cr
                                       4/(\eta+\eta_{eq}), &if
                                       $\eta>\eta_{eq}$.\cr}
\label{e6}
\end{equation}
Once again it is straightforward to solve Eq.~$(\ref{e4})$ giving
\begin{equation}
h_s(\eta) = \cases{j_0(k\eta), &if $\eta<\eta_{eq}$;\cr
                 (\eta_{eq}/\eta)\left(A_s j_1(k\eta) +
                 B_s y_1(k\eta)\right), &if $\eta>\eta_{eq}$.\cr}
\label{e7}
\end{equation}
The subscript $s$ denotes the fact that we are using the sudden
approximation~$(\ref{e6})$ . Notice that $h_s(\eta)$
satisfies the boundary conditions~$(\ref{e5})$, while the
integration constants $A_s$ and $B_s$ are determined by requiring
$h(\eta)$ and $dh/d\eta$ be continuous at $\eta=\eta_{eq}$. This gives
\begin{eqnarray}
A_s &=& 4\left(\sin k\eta_{eq} + \frac{\cos k\eta_{eq}}{2k\eta_{eq}}
        + \frac{\sin k\eta_{eq}}{4(k\eta_{eq})^2} -
        \frac{\sin^3k\eta_{eq}}{2(k\eta_{eq})^2} \right),
\nonumber \\
B_s &=& -4\left(\cos k\eta_{eq} - \frac{\sin k\eta_{eq}}{2k\eta_{eq}}
        - \frac{\cos k\eta_{eq}}{2(k\eta_{eq})^2} +
        \frac{\cos^3k\eta_{eq}}{2(k\eta_{eq})^2} \right).
\label{e8}
\end{eqnarray}
Notice also that $h_s\to1$ as $k\eta\to0$, and by this criterion
$h_s$ is a scale-invariant solution. Similar solutions can also be
found in Ref.~\cite{doro}.

Of course, the evolution of the universe is not nearly this sudden.
There is instead a gradual change in the equation of state from the
early RD era into the MD universe of the present day with the scale
factor
\begin{equation}
a(\eta)=a_{eq}\eta\left(\eta+2\right)\;.
\label{e3}
\end{equation}
$a_{eq}$ is the scale
factor at $\eta=\eta_{eq}$. (Note that $\eta_{eq}\approx 0.41$ and the present
time $\eta_0\approx 76.5$ if we take $a_0/a_{eq}= 6000$.) Then
Eq.~$(\ref{e4})$ reduces to
\begin{equation}
\frac{d^2h}{d\eta^2} + \frac{4(\eta+1)}{\eta(\eta+2)}
\frac{dh}{d\eta} + k^2h =0.
\label{e9}
\end{equation}
This equation can be solved exactly in terms of spheroidal wave functions
\cite{erd,sah}. Doing so will not be very illuminating and we shall
instead use the WKB approximation to obtain an approximate analytical
expression.

First, we define
\begin{equation}
h(\eta) = \frac{y(\eta)}{\eta(\eta+2)}.
\label{e10}
\end{equation}
Then Eq.~$(\ref{e9})$ becomes
\begin{equation}
\frac{d^2y}{d\eta^2} + \left(k^2 - \frac{2}{\eta(\eta+2)}\right)y = 0,
\label{e11}
\end{equation}
which is analogous to the Schr\"oedinger equation for the
``wavefunction'' of the particle with energy $k^2$ in a ``radial''
potential $V(\eta) = {2}/\{\eta(\eta+2)\}$
since $\eta\ge 0$. Because the boundary conditions for $h(\eta)$ are given
at $\eta=0$, the problem now reduces to the quantum mechanical problem
of a particle with energy $k^2$ tunneling out of the potential
$V(\eta)$.

Notice that in this language the sudden approximation
corresponds to a particle tunneling out of the potential
\begin{equation}
V_s(\eta) = \cases{0, &if $\eta<\eta_{eq}$;\cr
            2/(\eta+\eta_{eq})^2&if $\eta>\eta_{eq}$.\cr}
\label{e13}
\end{equation}
The singularity in $V(\eta)$ at $\eta=0$ has been removed and has
been replaced by a potential well with width $\eta_{eq}$ and height $1/2$.
This would explain the resonances in the amplitude of $h_s(\eta)$ when
$k\eta_{eq} \sim n\pi$ for some integer $n$, as can be seen explicitly in
Eq.~$(\ref{e8})$.

Since $\eta\ge0$, to apply the WKB approximation define $\eta = e^x$
where $x\in (-\infty,\infty)$. Then, taking $y(x) = e^{x/2} u(x)$,
\begin{equation}
\frac{d^2u}{dx^2} + \left(e^{2x} k^2 - \frac{2e^x}{e^x+2} -
\frac{1}{4}\right)u =0.
\label{e14}
\end{equation}
{}From the standard WKB connection formulas across the turning
point, we find that for $\eta<\eta_T$,
\begin{eqnarray}
h_w(\eta) &=&\frac{A(k)}{\eta^{1/2}(\eta+2)} \frac{1}{\sqrt {\Gamma(k\eta)}}
                \exp\left[-\int^{k\eta_T}_{k\eta}\Gamma(s)\frac{ds}{s}\right],
\;\;{\rm and} \nonumber \\
\Gamma(s) &=& \left({1\over4}+{2s\over s+2k} - s^2 \right)^{1/2},
\label{e16}
\end{eqnarray}
while for $\eta>\eta_T$,
\begin{eqnarray}
h_w(\eta) &=&\frac{2A(k)}{\eta^{1/2}(\eta+2)} \frac{1}{\sqrt {K(k\eta)}}
           \cos\left[\int^{k\eta}_{k\eta_T}
           K(s)\frac{ds}{s}-{\pi\over 4}\right],
\;\;{\rm and} \nonumber \\
K(s) &=& \left(s^2- {1\over4}-{2s\over s+2k}\right)^{1/2}.
\label{e17}
\end{eqnarray}
The tunneling amplitude $A(k)$ is
\begin{equation}
A(k) = \sqrt{2\eta_T}
\exp\left[\int_0^{k\eta_T}\left(\Gamma(s)
-{1\over 2}\right) \frac{ds}{s}\right],
\label{e18}
\end{equation}
and $\eta_T$ is the turning point of the potential in Eq.~$(\ref{e14})$
which is determined by
\begin{equation}
(k\eta_T)^2 - {2\eta_T\over \eta_T+2} - {1\over4} = 0.
\label{e19}
\end{equation}
One can show that $1/2\le k\eta_T\le 3/2$.
We should also note that when $k\eta$ is large, and when $\eta\gg2$,
the transfer function $T(k)$ of Ref.~\cite{TWL} can be easily obtained
from the WKB solution: $T(k) \approx (2/3)k^{3/2}A(k)$.

We shall now compare these three approximate solutions with each other.
Taking the small $k\eta<<1$ limit corresponding to modes well outside
the horizon, we find that $h_{rm}(\eta)\approx h_s(\eta)\approx 1$
and these solutions are scale invariant. For the WKB solution, on the
other hand, we find that $h_w(\eta)\approx H(\eta)$,
an $\eta$ dependent function that is close to unity for all $\eta$.
This unexpected $\eta$-dependence is mainly due to the breakdown of
WKB approximation in this limit. Since $h$ satisfies the boundary
conditions~$(\ref{e5})$, despite this small deviation from
unity the wave amplitude will remain constant when it is outside the
horizon.

Next, the asymptotic large $\eta\gg1$ limit for $k$ fixed at some
finite value gives
\begin{equation}
h_{rm}(\eta)\approx -\frac{3}{(k\eta)^2}\cos{k\eta},
\label{e22}
\end{equation}
for the MD equation of state while for the sudden
approximation,
\begin{equation}
h_s(\eta)\approx\cases{-3\cos k\eta/(k\eta)^2 &if $k\ll1$;\cr
             4k\eta_{eq}\cos(k\eta-\pi/2-k\eta_{eq})/(k\eta)^2 &if
             $k\gg1$.\cr}
\label{e23}
\end{equation}
The WKB approximation gives
\begin{equation}
h_w(\eta)\approx\cases{{8\sqrt3\over 5}\cos(k\eta-\pi)/(k\eta)^2 &if
        $k\ll1$;\cr
          2\sqrt{2\over e}\cos(k\eta-\pi/4)/(k\eta^2) &if
          $k\gg1$.\cr}
\label{e24}
\end{equation}
For small $k$ all three results reduce to
Eq.~$(\ref{e22})$, except a slightly different amplitude and an extra phase
of $\pi$ in $h_w$. This is to be expected. When the wavelength of the
GW is much greater than $\eta_{eq}$, it does not ``see'' the RD era
and is affected only by the MD universe. Consequently, the SI
solution $h_{rm}$ in the MD era works very well for GW of
wavelength in the regime $k\eta_{eq}\ll1$ at late times.

The situation changes dramatically for short-wavelength gravity
waves, however. Denoting the amplitudes of $h_{rm}(\eta)$, $h_s(\eta)$
and $h_w(\eta)$ by $M_{rm}$, $M_s$ and $M_w$ respectively,
\begin{equation}
\frac{M_{rm}}{M_s}\to \frac{3}{4k\eta_{eq}}\qquad,\qquad
\frac{M_{rm}}{M_w}\to \frac{3}{2k}\sqrt{e\over 2}\qquad,\qquad
\frac{M_s}{M_w}\to 2\eta_{eq}\sqrt{e\over 2}
\label{e25}
\end{equation}
for $k\gg1/4$.
Since the WKB solution is a closer approximation of the actual
history of the universe, we see that the sudden approximation
is a very good approximation of the amplitude of the GW. The SI
solution in the MD era, on the other hand, severely {\it
underestimates\/} the amplitude of the GW when $k>1.75$.

Although the change in the amplitude is most drastic for large
$k$, the relative change in the phase of $h(\eta)$
between the SI solution and the sudden and WKB approximations are
present at {\it all\/} $k$ and may be significant \cite{gris2}. In
fact, we see that although the amplitude of $h_s$ very closely
approximates that of $h_w$, its phase is $k$ dependent which, more
importantly, does not approach any fixed value for large $k$.
Thus, for different values of $k$ the phase difference between $h_s$
and $h_w$ will differ. These phase differences can be seen explicitly
by comparing Eqs.~$(\ref{e22})$-$(\ref{e24})$ and also in {\bf
Fig.~1}. Here we have also plotted the numerical solution $h_n(\eta)$
to Eq.~$(\ref{e9})$.

In our sudden approximation we could have also approximated the
equation of state with an $a$ which has a discontinuous first
derivative; namely $da/d\eta=2a/\eta$ for $\eta>\eta_{eq}$. In this
case we would have found an $h_s$ which, for $k\gg4$, underestimates
the amplitude of the GW by a factor of $4$. Its phase difference,
however, approaches a constant value of $\pi/4$ at large $k$.

We shall now use these results to calculate the anisotropy of CMB
induced by the SI spectrum of GW as well as its spectral energy density. From
the Sachs-Wolfe effect, the formula for the power spectrum $C_l$ is given
in, for example, \cite{krau}.
Integrating from the decoupling time $\eta_{dec}\approx 1.54$ for
$a_0/a_{dec}=1100$ and taking the inflation parameter
$v=V_0/m_{Pl}^4$, where $V_0$ and $m_{Pl}$ is the de Sitter vacuum
energy and Planck mass respectively, we have numerically evaluated
$C_l$ by using both the SI solution $h_{rm}$ and the numerical
solution $h_n$. A comparsion of the two results
is given in {\bf Table~1}. As expected, for
small $l$ the differences are insignificant since at these values the dominant
contribution to $C_l$ come from small $k$ modes, precisely where the SI
solutions work well. Due to the difference in the factor of $1/k$ in the
amplitudes of the SI and WKB solutions there is a change in $C_l$ for
large $l$, but unlike the results from using transfer functions
\cite{TWL}, we find that the SI solution {\it overestimates\/} the
numerical solution around $l\sim100$. This, we believe, is due to the
phase shift between $h_n$ and $h_{rm}$ which cannot be taken into
account by using a simple transfer function.

As for the spectral energy density of the GW
\begin{equation}
\Omega_g\equiv \sum_{\lambda=+,\times} \frac{k}{\rho_c}
\frac{d\rho_\lambda}{dk},
\label{e28}
\end{equation}
which is a measure of the total amount of energy deposited into the
experiment at any frequency $k$. $\rho_c$ is the closure density of
the universe. To estimate the amplitude of this quantity, one
typically makes use of the following physical argument. Before
the GW enters the horizon, it is a pure scale-invariant wave with
$h=1$. After it enters the horizon, on the other hand, the GW behaves
effectively as radiation and will scale with $a$ as such. This gives
a $\Omega_g\sim10^{-13}$ for $v\sim10^{-9}$ \cite{all,krau}.

We can also calculate $\Omega_g$ directly, however. Since
\begin{equation}
\sum_{\lambda=+,\times} k\frac{d\rho_\lambda}{dk}
=\frac{v}{3\pi^2 G} k_{\rm phys}^2 \vert h(\eta)\vert^2,
\label{e28.1}
\end{equation}
where $G$ and $k_{\rm phys}$ are the gravitational constant and physical
wavenumber respectively, once $h(\eta)$ is known,
one can simply evaluate $\Omega_g(\eta)$ at the present time $\eta_0$.
In fact, using Eq.~$(\ref{e3})$, we find
that
\begin{equation}
\Omega_g(\eta_0)=\frac{2v}{9\pi}k^2 \vert h(\eta_0)\vert^2
\left(\frac{\eta_0^2+2\eta_0}{\eta_0+1}\right)^2.
\label{e29}
\end{equation}
Since $\eta_0\gg1$ and since any experimentally detectable $k$ is
extremely large in our units, we find that by using the SI solutions
in Eq.~$(\ref{e29})$,
\begin{equation}
{\Omega_g(\eta_0)}_{rm}=\frac{2v}{\pi}{1\over (k\eta_0)^2}\cos^2(k\eta_0).
\end{equation}
Because $k\sim 10^{9}$, the
amplitude of ${\Omega_g(\eta_0)}_{rm}$ is infinitesimally small,
contradicting the above physical argument. If, on the other hand,
we use either sudden approximation solution or the WKB solution, then
\begin{eqnarray}
{\Omega_g(\eta_0)}_{s}&=&\frac{32v}{9\pi}\left(\eta_{eq}\over\eta_0\right)^2
                        \cos^2(k\eta_0-\pi/2-k\eta_{eq}),\nonumber \\
{\Omega_g(\eta_0)}_{w}&=&\frac{16ve^{-1}}{9\pi}{1\over\eta_0^2}
                        \cos^2(k\eta_0-\pi/4),
\label{e31}
\end{eqnarray}
and the amplitude of both $\Omega_g$ is a constant independent of $k$.

Since $\Omega_g(\eta)$ is an oscillatory function, we have plotted in
{\bf Fig. 2} the {\it time average\/}
\begin{equation}
\overline\Omega_g \equiv {k\over
2\pi}\int^{\eta_0}_{\eta_0-2\pi/k} \Omega_g(\eta)d\eta,
\end{equation}
of $\Omega_g$ over a complete period at the present time verses $k$
for $k$ much larger than the horizon size $k_H=2\pi/\eta_0$ using the
four solutions of $(\ref{e4})$. It is,
moreover, this time average which is measured in experiment and not
$\Omega_g(\eta)$. For $k$ near $k_H$, on the other hand, we have
plotted $\Omega_g(\eta_0)$ itself evaluated at the
present time $\eta_0$ verses $k$ in {\bf Fig.~3} since for these
values of $k$, $\Omega_g$ oscillates too
slowly for the averaging to be meaningful. Once again we can see from
{\bf Fig.~2} that the ${\overline\Omega_g}_{rm}$ obtained using
$h_{rm}$ severely underestimates the actual ${\overline\Omega_g}_n$
obtained using $h_n$ for large $k$. As expected,
${\overline\Omega_g}_s$ and ${\overline\Omega_g}_w$ obtained using
$h_s$ and $h_w$ are both fairly good
approximations of ${\overline\Omega_g}_n$. Notice also the dip in
$\Omega_g(\eta_0)$ in $\bf Fig.~3$ when $k\approx 0.7k_H$. This sharp
decrease in $\Omega_g$ at the present horizon size is due the finite
size of the universe.

In summary, we have studied the temporal dispersion of a SI spectrum
of primordial GW. We have found that the standard SI solution used in
the literature severely underestimates the size of the short wavelength,
scale-invariant GW at the present time. Although we would thus expect
the $C_l$ calculated from the SI solution to always underestimate the
actual $C_l$ calculated numerically, this does not turn out to be the
case. For $l\sim100$, it will overestimate the actual $C_l$
and is due to the relative phase shift in the $h_n$ verses $h_{rm}$.
This underestimation by the SI solution also occurs with the
calculation of $\Omega_g$, which up to now has been calculated using
either indirect physical arguments or by using transfer function
methods \cite{TWL}. This problem does not exist when using either
the sudden approximation or the WKB solutions, however, and
physically reasonable results for $\Omega_g$ can be obtained.

We also see from our solutions that the GW picks up a shift in phase
as the universe undergoes the radiation to matter phase transition
which is completely overlooked not only in the standard SI solutions,
but also in the transfer function method. This phase shift of the GW
may conceivably be an
invaluable probe of the early universe before the hydrogen
recombination. We can imagine future experiments performing
interferometry on two incident GW coming from different parts of the sky with
their phase shifts being dependent not only on the coincident wavenumber $k$,
but also on the detailed structure of the radiation-matter phase transition.
More importantly, it can also serve as a probe of the anisotropy of the early
universe as, conceivably, one part of the universe could have
undergone the phase transition before another part, thereby producing
a relative phase shift. However, the presence of initial randomly
distributed phases will complicate this situation, as they should
appear in the form of random noise in the measurement of the phase
shift. Such experiments would require much greater
sensitivity or new techniques than is now possible, of course, and
would lie in the far future.

\begin{center}
{\bf Acknowledgements}
\end{center}

This work was supported in part by the R.O.C. NSC Grant Nos.
NSC82-0208-M-001-131-T and NSC83-0208-M001-69.

\newpage
\centerline{\bf Captions}
\vskip1.75truecm
\noindent{{\bf Table 1}. Table of $C_l$ values calculated with the
standard SI solutions, denoted by ${C_l}_{rm}$, and the numerical
solution of Eq.~$(\ref{e9})$, denoted by $C_l$.
\vskip1.75truecm
\noindent{{\bf Figure 1}. Graphs of $h_{rm}(\eta)$, $h_s(\eta)$,
$h_w(\eta)$ and $h_n(\eta)$ to
Eq.~$(\ref{e9})$ for $k=200k_H$ where $k_H=2\pi/\eta_0$. Notice that
among the three approximate solutions, only the WKB solution
accurately reproduces the phase of $h_n$.
\vskip1.75truecm
\noindent{{\bf Figure 2}. Graphs of ${\overline\Omega_g}_{rm}$,
${\overline\Omega_g}_s$, ${\overline\Omega_g}_w$ as well as
${\overline\Omega_g}_n$ verses $k$ for large $k$. Notice the drastic
drop in ${\overline\Omega_g}_{rm}$.
\vskip1.75truecm
\noindent{{\bf Figure 3}. Graphs of ${\Omega_g(\eta_0)}_{rm}$,
${\Omega_g(\eta_0)}_s$, ${\Omega_g(\eta_0)}_w$ as well as
${\Omega_g(\eta_0)}_n$ verses $k$ for small $k$. All three
approximate analytical solutions are good approximations of $h_n$ in
this regime. The dip in $\Omega_g(\eta_0)$ at $k\approx0.7k_H$ is
actually a point at which $\Omega_g$ vanishes and is due to the
current finite size of the universe.}
\newpage

{\bf Table 1}
$$
\vbox{\offinterlineskip
\hrule
\halign{&\vrule#&\strut$\>$#$\>$&\vrule#&\strut$\>$#$\>$&\vrule#
&\strut$\>$#$\>$\cr
height3pt&\omit&&\omit&&\omit&\cr
&\hfil $l$\hfil&&\hfil ${C_l}_{rm}$\hfil&&\hfil $C_l$\hfil&
&\hfil $C_l/{C_l}_{rm}$\hfil&\cr
height3pt&\omit&&\omit&&\omit&&\omit&\cr
\noalign{\hrule}
height8pt&\omit&&\omit&&\omit&&\omit&\cr
&\hfil 2  \hfil&&\hfil 7.74    \hfil&&\hfil 7.76   \hfil&&\hfil 1.003\hfil&\cr
&\hfil 20 \hfil&&\hfil 0.714   \hfil&&\hfil 0.718  \hfil&&\hfil 1.006\hfil&\cr
&\hfil 50 \hfil&&\hfil 0.255   \hfil&&\hfil 0.238  \hfil&&\hfil 0.93 \hfil&\cr
&\hfil 100\hfil&&\hfil 0.0611  \hfil&&\hfil 0.0383 \hfil&&\hfil 0.63 \hfil&\cr
&\hfil 150\hfil&&\hfil 0.00892 \hfil&&\hfil 0.00223\hfil&&\hfil 0.25 \hfil&\cr
&\hfil 200\hfil&&\hfil 0.000468\hfil&&\hfil 0.00180\hfil&&\hfil 3.85 \hfil&\cr
&\hfil 250\hfil&&\hfil 0.000479\hfil&&\hfil 0.00113\hfil&&\hfil 2.36 \hfil&\cr
height8pt&\omit&&\omit&&\omit&&\omit&\cr}
\hrule}
$$

\end{document}